# ELECTRIC CONDUCTIVITY OF LEAD IRON NIOBATE

K. ZIELENIEC, M. MILATA, and K. WÓJCIK

*Institute of Physics, University of Silesia,*

*ul. Uniwersytecka 4,*

*PL-40-007 Katowice, Poland*

INTRODUCTION

Compounds with perovskite type structure can be treated as model materials because of the richness of physical properties compared to the relative simply structure. Moreover, these materials have found many technical applications as e. g. Computer memories, pyroelectric sensors, piezoelectric transducers, and high density capacitors [1]. Lead iron niobate belongs to a large family of complex perovskite type ferroelectric oxides corresponding to the general formula $A(B^I B^{II})O_3$, where $B^I$ and $B^{II}$ ions are randomly distributed in the octahedral positions. At room temperature PFN exhibits ferroelectric and antiferromagnetic properties. The phase transition from rhombohedral (ferroelectric) to cubic (paraelectric) phase takes place at 385K or 388K [2 - 5]. According to [6] the transition from antiferromagnetic phase to paramagnetic one takes place at 143K (magnetic Néel temperature). Alike another disordered complex perovskites the PFN show a relaxor – type dielectric behaviour. The measurements of the dielectric parameters are rather difficult because of a high electrical conductivity $\sigma$ of this material [7]. High conductivity of the PFN samples can be diminished, by doping with small amount of lithium [7, 8]. The aim of this work was to examine temperature dependencies of the d. c. conductivity $\sigma$ and the Seebeck coefficient $\alpha$ for PFN "pure" and PFN doped with lithium.

EXPERIMENTAL

*Preparation of samples*

Ceramic samples were obtained by sintering analytically pure oxides: $PbO$, $Nb_2O_5$, and $Fe_2O_3$. This was performed in three stages:

- calcination of the $PbO – Nb_2O_5 – Fe_2O_3$ mixture, weighed according to formula $PbFe_{1/2}Nb_{1/2}O_3$, at $900^0C$ for 3 hours
- sintering of $PbFe_{1/2}Nb_{1/2}O_3$ at $1050^0C$ for 3 hours
- final sintering in a protective atmosphere $(PbO + PbO_2 + ZrO_2)$ at $1160^0C$ for 4 hours.

Samples were sintered in alumina crucibles. The final sintering was performed by a method similar to that described by Snow [9]. In the case of doped materials 0.25 mol%, 0.5 mol%, and 1.0 mol% $Li_2CO_3$ were added to the $PbO – Nb_2O_5 – Fe_2O_3$ mixture before the calcination process. The samples showed minor porosity and good mechanical properties. The colour of samples was dark brown with the rusty shade.

Single crystals were obtained by the "flux" method from $PbFe_{1/2}Nb_{1/2}O_3 – PbO – B_2O_3$ solution. This mixture was homogenised in a sealed platinum crucible at soaking temperature $T_S = 1200^0C$ for 12 hours. The temperature was then lowered to the final temperature $T_F = 820^0C$ with cooling rate $dT/dt$ = 4K/hour. After separation of the solvent at temperature $T_F$, crystals were cooled to room temperature in the furnace. The remaining solvent was etched using a boiling acetic acid solution. Rectangular prism – shaped single crystals with edges up to 4 mm of dark brown (nearly black) colour were obtained. The X – ray analysis showed the perovskite structure of both ceramic samples and single crystals. Results of the X - ray investigations will be published separately. Plate – shaped ceramic or single crystal samples with dimensions about $4 \times 4 \times 1 mm^3$ were used for measurements. Electrodes were deposited on the sample using silver paste.

*Electric conductivity, and Seebeck coefficient.*

Electric conductivity was measured in the temperature range from 20 to $450^0C$ using an electrometer and a recorder. The voltage 0.5V was applied to the sample. The Seebeck coefficient was measured in the temperature range from 100 to $450^0C$. The sample was placed in a thermostat between heated silver blocks, which make possible to create a temperature gradient inside the sample.

The temperature difference $\Delta T$, at each average temperature, was changed from 0 to 10K. Values of Seebeck coefficient $\alpha$ were determined from the slope of the thermal electromotive force $E_T$ linear dependencies on the temperature difference $\Delta T$.

The temperature dependencies of electric conductivity for ceramic samples and a single crystal are shown on Figure 1.

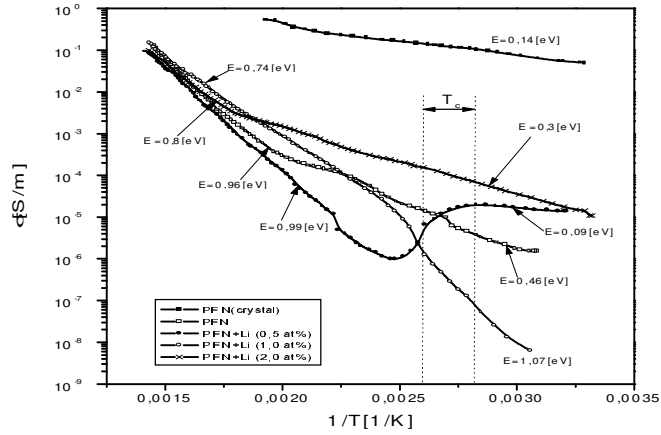

Figure 1. Temperature dependencies of electric conductivity (cooling process). $E$ – values of activation energy calculated from Arrhenius equation $\sigma = \sigma_0 e^{-\frac{E}{kT}}$.

The single crystal exhibits the highest value of conductivity. The conductivity of ceramic samples is markedly lower than that observed in the single crystals. The values of conductivity obtained for ceramic samples, markedly differs within temperature range from the room temperature to about $300^0$C. It is clearly visible that, in this temperature range, the conductivity is markedly dependent on the concentration of the Li admixture. Values of conductivity are similar above $300^0$C. Electric conductivity, in some ranges of temperature obeys the Arrhenius law

$$\sigma = \sigma_0 e^{-\frac{E}{kT}}.$$

Values of activation energy, calculated from this law, are marked on the plots $\ln\sigma = f\left(\frac{1}{T}\right)$ (Fig.1). The temperature dependencies of Seebeck coefficient are presented on Figure 2.

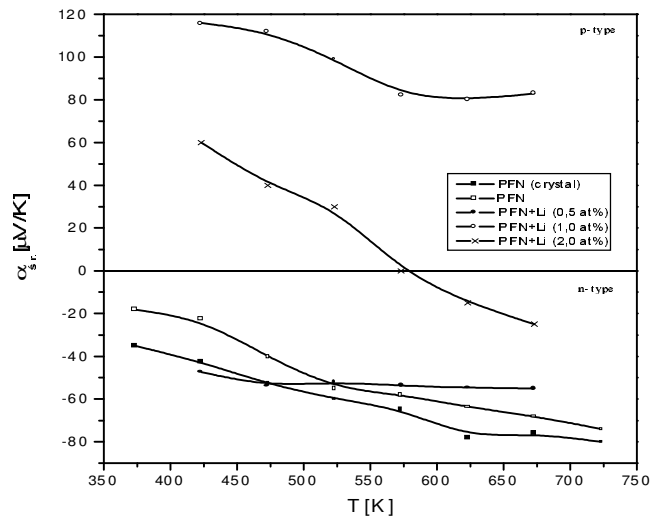

Figure 2. Seebeck coefficient versus average temperature for the single crystal, and ceramic samples.

The single crystal and ceramic samples PFN, and PFN + 0.5 at % exhibit n – type conductivity. The conductivity of p – type was observed for PFN + 1.0 at % Li.

Samples PFN + 1.0 at % Li are p – type semiconductors up to $300^0$C, and n – type semiconductors above this temperature. It is note worthy that the $\sigma(T)$ dependence obtained for PFN + 0.5 at % Li, shows an anomaly (a part with the negative temperature coefficient of conductivity) near the phase transition temperature reported in papers [2 – 5]. The change in the slope of the graph $\ln\sigma = f\left(\dfrac{1}{T}\right)$ obtained for PFN + 1.0 at % Li samples seems to be associated with this phase transition too. In the case of remaining samples there are no anomalies, which could be associated with the phase transition. Moreover, it should be noted that the charge carrier activation energy value for PFN + 0.5 at % Li is very small ($E = 0.09$eV) up to about $100^0$C.

DISCUSSION

Both single crystals and undoped ceramics samples exhibit n – type conductivity, which may be determined by the oxygen deficiency. Nonstoichiometry in the oxygen sublattice, can be a result of reduction processes proceeding at high temperature, during technological process. This reduction process may be written as follows:

$$Pb_2NbFeO_6 \rightarrow Pb_2NbFeO_{6-x} + \frac{1}{2}xO_2$$

Because of the oxygen deficiency oxygen vacancies ($V_O$) appears in the crystal lattice. These vacancies are donor centres because of ionisation processes, which may be written as follows:

$$V_O \Leftrightarrow V_O^\bullet + e^-$$
$$V_O^\bullet \Leftrightarrow V_O^{\bullet\bullet} + e^-$$

The reduction processes may lead to the creation of oxygen vacancies or valency change in *Nb* and *Fe* ions. It may be written as follows:

$$V_O + Nb_{Nb} \Leftrightarrow V_O^\bullet + Nb'_{Nb} \quad \text{or} \quad V_O + Nb_{Nb}^{5+} \Leftrightarrow V_O^\bullet + Nb_{Nb}^{4+}$$

$$V_O + Fe_{Fe} \Leftrightarrow V_O^\bullet + Fe'_{Fe} \quad \text{or} \quad V_O + Fe_{Fe}^{3+} = V_O^\bullet + Fe_{Fe}^{2+}$$

In that way the lattice defects such as $V_O$, $V_O^\bullet$, $F'_{Fe}$, $Nb'_{Nb}$, and complexes: $V_O^\bullet Nb'_{Nb}$, and $V_O^\bullet Fe'_{Fe}$ can be donor centres. Activation energy of the conductivity is rather small (0.14eV for single crystal, and 0.46 – 0,80eV for ceramic samples). Hence electric conduction is extrinsic conduction and its activation energy is believed to be, mainly connected with charge carriers mobility. It shows on a hopping mechanism of the conductivity associated with the occurence of oxygen vacancies and valency change of the transition elements *Fe* and *Nb*. This conductivity mechanism was described, among others, by Suchet [10].

Relatively large n – type conductivity, associated with the nonstoichiometry, can be compensated by doping with cations of a valency lower than that of the host ions. The impurity ions may replace *Pb*, *Nb*, and *Fe* ions, creating acceptor centres. Lithium as an acceptor impurity can be used, in the case of the PFN samples. The doping with 1.0 at % Li cause a marked decrease in electric conductivity, in the temperature range from room temperature to $120^0$C. Moreover it brings about the change of the type of electric conductance from n to p – type. However further increasing in the contents of the impurity cause an increase in conductivity. It seems probably, that lithium atoms (ions) can replace the host ions $(Li'_{Pb}, Li''_{Fe})$ or occupy the interstitial positions ($Li_i$). The replacement of iron ions by lithium may be expressed as follows:

$$Pb_2NbFeO_6 + \frac{1}{2}xLi_2O + \frac{1}{2}xO_2 \rightarrow Pb_2NbFe_{1-x}Li_xO_6 + \frac{1}{2}xFe_2O_3$$

or
$$Li_2O + \frac{1}{2}O_2 \rightarrow 2Li''_{Fe} + 2Pb^{\bullet\bullet}_{Pb} + 2O_O.$$

Last reaction may be presented in another way:

$$Li_2O + \frac{1}{2}O_2 \rightarrow 2Li''_{Fe} + 4h^{\bullet} + O_O$$

$$[Li''_{Fe}] = [Pb^{\bullet\bullet}_{Pb}] \quad or \quad 2[Li''_{Fe}] = [h^{\bullet}] \quad \text{the charge neutrality condition.}$$

$Li^{1+}$ cation, replacing $Fe^{3+}$ ion, create an acceptor centre, which negative electric charge can be compensate by valency change in *Pb* ion (controlled electron disorder) [11]. In this way the PFN becomes a p – type semiconductor. The doping with lithium can therefore compensate the n – type conductivity appearing in nonstoichiometric (oxygen deficient PFN). If lithium atoms occupy interstitial positions they will be donor centres. Their ionisation can be presented as follows:

$$Li_{Li} \Leftrightarrow Li^{\bullet} + e^{-}$$

Results of measurements show, that lithium ions (atoms) can replace host cations (e. g. *Fe*) or they can occupy the interstitial positions in the crystal lattice. The variety of defects appearing in material examined as result of nonstoichiometry or doping allows to treat it as heavy doped and partly compensated semiconductor. The change in concentration of impurities in such material not only changes degree of compensation but it can change a mechanism of conductance as well. The hopping motion of charge carriers may be observed [12]. It seems probably that in the samples of PFN + 1.0 at % Li the hopping motion in the impurity band is a dominant mechanism of conductivity in relatively low temperatures. Small value of activation energy (0.09eV) seems to confirm this fact.

SUMMARY

The lead iron niobate (PFN) samples exhibit n – type conductance. This conductance can be brought about nonstoichiometry connected with the oxygen deficiency. Relatively large conductivity of the n – type can be compensated (and diminished) by doping with lithium as an acceptor impurity. The contents of lithium of order 1.0 at % cause a marked decrease in electric conductivity in the low temperature range, and the alteration of the electric conductance from n to p – type. However the increase in the concentration of lithium (up to 2.0 at %) leads to the new increase in conductivity.

Moreover the change of the electric conductance type from p to n has been found. These phenomena may be probably caused by various possibilities of Li atoms incorporating into PFN lattice.

Large values of electric conductivity, and its heavy dependence on the nonstoichiometry, and a contents of the impurity, exert a marked influence on dielectric properties of PFN samples. In particular the behaviour of metal – semiconductor junctions is very important, because dielectric parameter measurements have usually been performed for samples with metal electrodes. However, results of dielectric parameter measurements in the "metal – PFN – metal" system will be a subject of separate work.